\documentclass[10pt,twocolumn,letterpaper]{article}

\usepackage{cvpr}              

\usepackage{graphicx}
\usepackage{booktabs,makecell}

\usepackage[dvipsnames]{xcolor}
\usepackage{color}
\usepackage{colortbl}
\usepackage{caption}
\usepackage{floatrow}
\usepackage{amsmath,bm, amssymb, amsfonts} 
\usepackage{nicefrac}
\usepackage{arydshln}

\usepackage{dsfont}
\usepackage{multirow}
\usepackage{subcaption}
\usepackage{xspace}
\usepackage{array}
\usepackage{textcomp}
\usepackage{tikz}
\usetikzlibrary{calc,trees,positioning,arrows,chains,shapes.geometric,patterns,%
	decorations.pathreplacing,decorations.pathmorphing,shapes,%
	matrix,shapes.symbols,plotmarks,decorations.markings,shadows}
\usepackage{array}
\usepackage{arydshln}
\usepackage{pifont}
\usepackage{etoolbox}

\usepackage{comment}
\usepackage{mathtools}

\usepackage{array}
\usepackage{diagbox}
\usepackage{listings}
\usepackage{algorithm}
\usepackage[pagebackref,breaklinks,colorlinks,citecolor=blue]{hyperref}

\usepackage{amsmath,amsfonts,bm}

\def\eqref#1{equation~\ref{#1}}

\def\1{\bm{1}}

\def\vs{{\bm{s}}}

\DeclareMathAlphabet{\mathsfit}{\encodingdefault}{\sfdefault}{m}{sl}
\SetMathAlphabet{\mathsfit}{bold}{\encodingdefault}{\sfdefault}{bx}{n}

\def\gG{{\mathcal{G}}}

\def\gM{{\mathcal{M}}}

\def\gQ{{\mathcal{Q}}}
\def\gR{{\mathcal{R}}}

\def\gT{{\mathcal{T}}}

\makeatletter
\DeclareRobustCommand\onedot{\futurelet\@let@token\@onedot}
\def\@onedot{\ifx\@let@token.\else.\null\fi\xspace}

\def\eg{\emph{e.g}\onedot} 
\def\ie{\emph{i.e}\onedot} 
 
\def\etc{\emph{etc}\onedot} \def\vs{\emph{vs}\onedot}

\makeatother

\newcommand{\tabincell}[2]{\begin{tabular}{@{}#1@{}}#2\end{tabular}}
\newcommand{\Th}[1]{\textsc{#1}}
\newcommand{\mr}[2]{\multirow{#1}{*}{#2}}
\newcommand{\mc}[2]{\multicolumn{#1}{c}{#2}}

\newcommand{\ltnorm}[1]{\left\|#1\right\|_2}

\newcommand{\bI}[1]{\color{OliveGreen}{($\uparrow$ \bf{#1})}}
\newcommand{\bIP}[1]{\color{OliveGreen}{($\uparrow$ #1)}}

\newcolumntype{H}{>{\setbox0=\hbox\bgroup}c<{\egroup}@{}}

\newcommand{\mypartight}[1]{\vspace{0cm}{\noindent\textbf{#1}}}

\newfloatcommand{capbfigbox}{figure}[][\FBwidth]
\newfloatcommand{hcapbfigbox}{figure}[][1\linewidth]

\usepackage{pgfplots, pgfplotstable}
\usepgfplotslibrary{groupplots}
\pgfplotsset{compat=newest}
\usepgfplotslibrary{fillbetween}

\newcommand{\oxf}{MidnightBlue}

\newcommand{\paris}{RedViolet}

\newcommand{\valc}{PineGreen}

\newcommand{\ourscolor}{Maroon}
\newcommand{\ourscolors}{NavyBlue}
\newcommand{\howcolor}{Black}
\newcommand{\csdcolor}{Black}

\newcommand{\oxfmark}{o}
\newcommand{\parmark}{triangle}
\newcommand{\valmark}{square}

\tikzset{every mark/.append style={solid}}
\pgfplotsset{
	grid=both, width=\linewidth, try min ticks=5,
    legend cell align=left, 
    legend style={fill opacity=0.8},
	ylabel near ticks,
    xlabel near ticks,
    every tick label/.append style={font=\footnotesize},
}

\pgfplotsset{
	oxmed/.style={thick, color=\oxf, mark=o},
    oxhard/.style={thick, dashed, color=\oxf, mark=o},
    pamed/.style={thick, color=\paris, mark=star}, 
    pahard/.style={thick, dashed, color=\paris, mark=star},
    val/.style={thick, color=\valc, mark=x},
    oursoxf/.style={thick, color=\ourscolor, mark=o},
    oursoxfs/.style={thick, dashed, color=\ourscolors, mark=triangle},
    howoxf/.style={thick, dashed, color=\howcolor, mark=\oxfmark},
    csdoxf/.style={thick, dotted, color=\csdcolor, mark=\valmark},
    ourspar/.style={thick, color=\ourscolor, mark=\parmark},
    howpar/.style={thick, dashed, color=\howcolor, mark=\parmark},
    oursval/.style={thick, color=\ourscolor, mark=\valmark},
    howval/.style={thick, dashed, color=\howcolor, mark=\valmark},
    numean/.style={thick, color=\ourscolor, mark=none},
    numin/.style={thick, color=gray, mark=none},
    wid/.style={thick, color=\ourscolor, mark=o, mark size=0.5pt},
    woid/.style={thick, densely dashdotted, color=RedOrange, mark=o, mark size=0.5pt},
}

\tikzstyle{lay} = [draw,minimum width=20pt,inner sep=2pt]
\tikzstyle{selfpre} = [draw,minimum width=70pt,inner sep=2pt]

\tikzstyle{inp} = [selfpre,fill=black!15]
\tikzstyle{gap} = [lay,fill=brown!25]

\tikzstyle{self} = [selfpre,fill=yellow!15]
\tikzstyle{addnorm} = [selfpre,fill=orange!25]

\tikzstyle{down} = [lay,fill=yellow!15,trapezium,trapezium angle=110]
\tikzstyle{up} = [lay,fill=yellow!15,trapezium,trapezium angle=70]
\tikzstyle{fix} = [lay,fill=blue!10]
\tikzstyle{cat} = [lay,fill=green!15]
\tikzstyle{fun} = [lay,ellipse,fill=orange!25]
\tikzstyle{outp} = [lay,fill=red!15]
\tikzstyle{dim} = [black!50]
\tikzstyle{key} = [red]
\tikzstyle{tight} = [inner sep=0pt,outer sep=0pt]
\tikzstyle{node}  = [draw,circle,tight,minimum size=12pt,anchor=center]
\tikzstyle{op}    = [draw,circle,tight]
\tikzstyle{dot}   = [fill,draw,circle,inner sep=1pt,outer sep=0]
\tikzstyle{pt}    = [fill,draw,circle,inner sep=1.5pt,outer sep=.2pt]
\tikzstyle{box}   = [draw,thick,rectangle,inner sep=3pt]
\tikzstyle{high}  = [black!60]
\tikzstyle{group} = [high,box,opacity=.5]
\tikzstyle{dim1}  = [fill opacity=.3,text opacity=1]
\tikzstyle{dim2}  = [fill opacity=.5,text opacity=1]
\tikzstyle{dim3}  = [fill opacity=.7,text opacity=1]
\tikzstyle{rectc} = [tight,transform shape]
\tikzstyle{rect}  = [rectc,anchor=south west]

\tikzstyle{outer} = [draw, minimum width=70pt,minimum height=70pt,dashed,fill opacity=.3,text opacity=1]

\makeatletter
\def\adl@drawiv#1#2#3{%
	\hskip.5\tabcolsep
	\xleaders#3{#2.5\@tempdimb #1{1}#2.5\@tempdimb}%
	#2\z@ plus1fil minus1fil\relax
	\hskip.5\tabcolsep}
\newcommand{\cdashlinelr}[1]{%
	\noalign{\vskip0.7\aboverulesep
		\global\let\@dashdrawstore\adl@draw
		\global\let\adl@draw\adl@drawiv}
	\cdashline{#1}
	\noalign{\global\let\adl@draw\@dashdrawstore
		\vskip0.7\belowrulesep}}
\makeatother

\def\cvprPaperID{1180} 
\def\confName{CVPR}
\def\confYear{2023}

\begin{document}
	
\title{Asymmetric Feature Fusion for Image Retrieval}

\author{Hui~Wu$^{1}$\quad Min~Wang$^{2*}$\quad Wengang~Zhou$^{1,2*}$\quad Zhenbo~Lu$^{2}$ \quad Houqiang~Li$^{1,2}$ 
	\\
	\normalsize $^{1}$CAS Key Laboratory of Technology in GIPAS, University of Science and Technology of China \\
	\normalsize $^{2}$Institute of Artificial Intelligence, Hefei Comprehensive National Science Center \\
	\tt\small wh241300@mail.ustc.edu.cn, \{wangmin,luzhenbo\}@iai.ustc.edu.cn, \{zhwg,lihq\}@ustc.edu.cn
}

\maketitle

\newcommand\blfootnote[1]{%
	\begingroup 
	\renewcommand\thefootnote{}\footnote{#1}%
	\addtocounter{footnote}{-1}%
	\endgroup 
}

\begin{abstract}
	In asymmetric retrieval systems, models with different capacities are deployed on platforms with different computational and storage resources. Despite the great progress, existing approaches still suffer from a dilemma between retrieval efficiency and asymmetric accuracy due to the limited capacity of the lightweight query model. In this work, we propose an \textbf{A}symmetric \textbf{F}eature \textbf{F}usion (AFF) paradigm, which advances existing asymmetric retrieval systems by considering the complementarity among different features just at the gallery side. Specifically, it first embeds each gallery image into various features, e.g., local features and global features. Then, a dynamic mixer is introduced to aggregate these features into compact embedding for efficient search. On the query side, only a single lightweight model is deployed for feature extraction. The query model and dynamic mixer are jointly trained by sharing a momentum-updated classifier. Notably, the proposed paradigm boosts the accuracy of asymmetric retrieval without introducing any extra overhead to the query side. Exhaustive experiments on various landmark retrieval datasets demonstrate the superiority of our paradigm.
\end{abstract}

\blfootnote{*Corresponding Author: Min~Wang and Wengang~Zhou.} 

\section{Introduction}
\label{sec:intro}

\begin{figure}[t]
	\centering
	\begin{subfigure}[b]{\linewidth}
		\centering
		\includegraphics[width=0.99\linewidth]{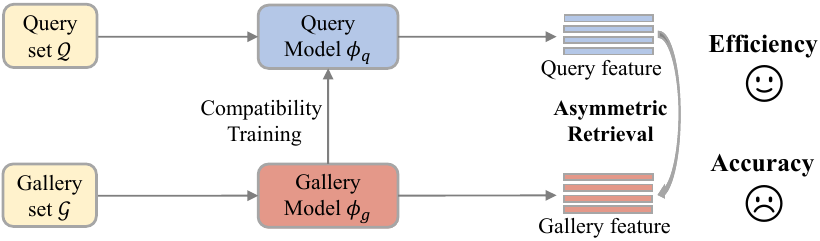}%
		\caption{Previous single feature pipeline~\cite{HVS,CSD,BCT,AML}.\vspace*{2pt}}
		\label{fig:BCT_pipline}
	\end{subfigure}
	\begin{subfigure}[b]{\linewidth}
		\centering
		\vspace*{5pt}
		\includegraphics[width=0.99\linewidth]{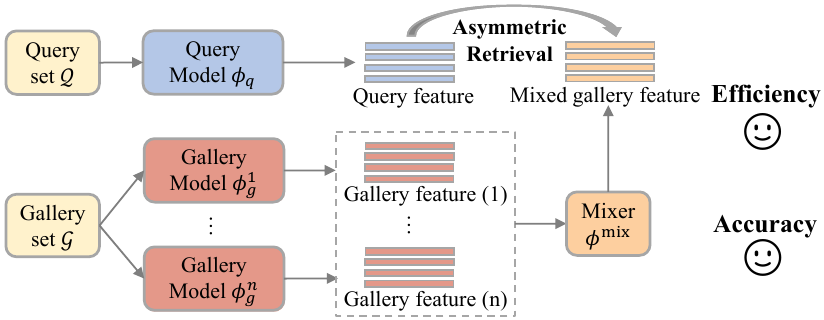}%
		\caption{Our asymmetric feature fusion pipeline.\vspace*{-2pt}}
		\label{fig:ours_pipline}
	\end{subfigure}
	\vspace*{-15pt}
	\caption{Illustration of (a) \textbf{previous single-feature asymmetric retrieval pipeline} and (b) our \textbf{asymmetric feature fusion paradigm}. Due to limited capacity of the lightweight model, existing pipeline achieves efficiency for the query side at the cost of retrieval accuracy degradation. In contrast, our approach enhances existing asymmetric retrieval pipeline from the perspective of \emph{gallery feature fusion}. For efficient retrieval, a dynamic mixer is introduced to aggregate multiple gallery features into a compact embedding. Query model and mixer are jointly trained with compatible constraints. Our method realizes high efficiency without sacrificing retrieval accuracy.\vspace*{-15pt}}
	\label{fig:pipline}
\end{figure}

\begin{figure}[t]
	\centering
	\vspace*{5pt}
	\resizebox{1.0\linewidth}{!}{
		\input{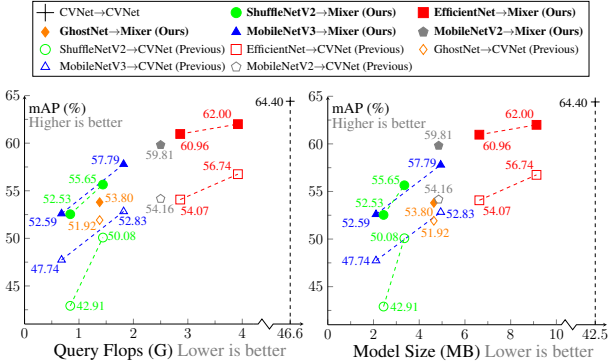}
	}
	\vspace*{-15pt}
	\caption{\textbf{Average mAP \vs FLOPs/Model Size of the query model} for $\gR$Oxf + 1M~\cite{roxpar} dataset. The notation format ``query model $\rightarrow$ gallery model'' in the legend means embedding queries with the query model and retrieving in a gallery set embedded by the gallery model. \textbf{A line connecting the dots with one color} represents \textbf{a family of lightweight models with different model sizes}. \emph{Previous}: The latest asymmetric retrieval method CSD~\cite{CSD} is adopted to train query model with CVNet~\cite{CVNet} deployed as gallery model. \emph{Ours}: our paradigm utilizes CVNet, Token~\cite{Token}, DELG~\cite{DELG} and DOLG~\cite{DOLG} to generate aggregated gallery features and trains the mixer and query model jointly.\vspace*{-15pt}}
	\label{fig:performace_compare}
\end{figure}

Image retrieval~\cite{videogoogle,nister2006scalable,hammingembeding,RMAC,gem,GLDv2} has been studied for a long time in the literature. Typically, high-performing image retrieval systems deploy a large powerful model to embed both query and gallery images, which is widely known as \textbf{symmetric image retrieval}. However, in some real-world applications, \eg, mobile search, the query side is constrained by resource limitation and thus cannot meet the overhead of deploying a large model. To this end, the paradigm of \textbf{asymmetric image retrieval} is first proposed in HVS~\cite{HVS} and AML~\cite{AML}, which has attracted increasing attention from the community~\cite{LCE,CSD,BCT,ijcai2022p225,suma2022large}. In such a paradigm, deep representation models with different capacities are first trained to be compatible and then deployed on platforms with different resources to strike a balance between retrieval accuracy and efficiency, as shown in Fig.~\ref{fig:BCT_pipline}.

For an asymmetric retrieval system, the most crucial thing is to ensure that the embedding spaces of different models are well aligned. To this end, BCT~\cite{BCT} first proposes a \emph{backward-compatible} learning framework, in which the classifier of the gallery model is inherited to guide the learning of the query model. Recently, various efforts have been devoted to improving cross-model feature compatibility in terms of training objectives~\cite{AML,CSD,LCE,NCCL,CLVS}, model structures~\cite{HVS,RFNS}, \etc. Despite the great progress, a dilemma is still unresolved, \ie, the accuracy of asymmetric retrieval is still unsatisfactory compared to that of symmetric retrieval, especially in limited-resource and large-scale scenarios, as shown in Fig.~\ref{fig:performace_compare} (${\color{red}{\square}},{\color{orange}{\Diamond}},\dots,\color{blue}{\triangle}$ \vs $+$). We argue that such dilemma is due to the low capacity of lightweight query model, which cannot perfectly achieve feature compatibility with the static powerful gallery model.

To alleviate above issue, we introduce a new paradigm named \textbf{A}symmetric \textbf{F}eature \textbf{F}usion (AFF). It boosts the accuracy of existing asymmetric retrieval systems by considering the complementarity among different features, as shown in Fig.~\ref{fig:ours_pipline}. On the gallery side, it deploys several large powerful models on the cloud to extract diverse features, \eg, local features, which are suitable for capturing local matches, or global features that are effective for holistic semantic matching. For efficient retrieval, a dynamic mixer is further proposed to aggregate diverse gallery features into compact embedding, which allows efficient vector search~\cite{PQ,OPQ} to be exploited. As for the query side, queries are embedded with a single lightweight model. It eliminates time-consuming multiple feature extraction and aggregation processes, realizing a solution suitable for resource-constrained platforms. During training, all the gallery models are fixed, while the mixer and query model are trained jointly by a momentum-updated classifier for achieving feature compatibility.

Compared to previous retrieval approaches, the proposed paradigm has two unique advantages. First, it fuses various features on the gallery side, which notably advances the retrieval accuracy of existing asymmetric retrieval systems. Although the extraction and aggregation processes increase the computational complexity, they are typically performed on resource-rich cloud platforms. In addition, gallery images are embedded offline in advance, whose computational overhead has no influence on the query side. Second, compared with multi-feature fusion methods, our paradigm only deploys a single lightweight model on the query side, which is free of the complex and time-consuming multi-feature extraction and aggregation. Thus, it introduces no extra computational and storage overhead for the query side. Overall, with the proposed asymmetric feature fusion paradigm, our approach achieves high retrieval efficiency and accuracy simultaneously, as shown in Fig.~\ref{fig:performace_compare} (${\color{red}{\square}}$ \vs ${\color{red}{\blacksquare}}$,$\dots$,$\color{blue}{\triangle}$ \vs $\color{blue}{\blacktriangle}$). To evaluate our approach, comprehensive experiments are conducted on popular landmark retrieval datasets. The proposed paradigm realizes promising performance improvement for existing asymmetric retrieval systems and leads to the state-of-the-art results across the public benchmarks. 
 
\section{Related Work}
\label{sec:related}

\noindent \textbf{Feature Representation}. In image retrieval, feature representation plays a key role. Hand-crafted local features~\cite{lowe2004distinctive,bay2006surf} are widely used in early image retrieval systems~\cite{videogoogle,oxford,zhou2010spatial}. Recently, local features extracted from convolutional neural networks (CNNs) are shown to be more effective~\cite{DELF,HardNet,SOSNet,Superpoint}. They learn feature detection and representation jointly by attention mechanism~\cite{DELF,HOW,MDA,Superfeature} or non-maximal suppression~\cite{D2Net}. The detected local features are further utilized for geometric verification~\cite{oxford} or aggregated into compact representations by VLAD~\cite{VLAD}, ASMK~\cite{ASMK}, \etc, for efficient retrieval. Recently, global features such as RMAC~\cite{RMAC}, GeM~\cite{gem}, DELG~\cite{DELG}, DOLG~\cite{DOLG}, Token~\cite{Token} and CVNet~\cite{CVNet}, are typically extracted from CNNs by spatial pooling~\cite{gem,DALG,GLAM}, which demonstrate more effectiveness in holistic semantic matching over local features.

Despite the great progress, existing image retrieval systems usually deploy large powerful models for high retrieval accuracy. However, some real-world applications need to deploy query models on resource-constrained platforms, \eg, mobile phones, which cannot meet the demand of large models for computational and storage resources. To address this issue, our approach focuses on the setting of asymmetric retrieval, where the query side deploys a lightweight model while the gallery side applies a large one. 

\noindent \textbf{Feature Compatibility}. 
The paradigm of \emph{feature compatibility learning} is first proposed by BCT~\cite{BCT}. It enforces the feature of the query model to be close to the corresponding class centroids of the gallery model. Under this paradigm, several efforts~\cite{AML,LCE,HVS,CSD,ijcai2022p225} have been devoted to improving the feature compatibility across different models. Specifically, AML~\cite{AML} introduces asymmetric regression loss and contrastive loss to train the query model. CSD~\cite{CSD} takes a step further by constraining the query model to maintain the nearest neighbor structure in the embedding space of the gallery model. Recently, LCE~\cite{LCE} proposes to align the classifiers of different models with a tight boundary loss. HVS~\cite{HVS} further resorts to neural architecture search technique to search for the optimal compatibility-aware model architecture. FCT~\cite{FCT} stores ``side information'', which is later leveraged to transfer the gallery features for other retrieval tasks. Besides, when solving the model regression problem, methods including PCT~\cite{PCT}, REG-NAS~\cite{RFNS} and RACT~\cite{RACT}, also utilize feature compatibility to alleviate ``negative flip''. 

Differently, to boost existing asymmetric retrieval systems, we introduce a new asymmetric feature fusion paradigm. It enhances the discriminativeness of image features by aggregating diverse features just at the gallery side. Our approach is readily combined with existing methods to achieve better asymmetric retrieval accuracy efficiently.

\noindent \textbf{Lightweight Network}. The architecture of deep convolutional neural networks~\cite{AlexNet,ResNet} has been evolving for many years. As the application complexity increases~\cite{Clip}, model size becomes larger, requiring more computational and storage resources. However, besides accuracy, resource overhead is another important consideration in real-world applications. Real-world tasks usually expect to deploy optimal models on the target resource-constrained platforms. The immediate demand motivates a series of work, \eg, MobileNets~\cite{MobileNets,Mobilenetv2}, ShuffleNets~\cite{shufflenet,shufflenetv2}, GhostNets~\cite{ghostnet} and EfficientNets~\cite{Efficientnet}, for lightweight model design. 

In this work, we focus on asymmetric retrieval in resource-constrained scenarios. Various lightweight models mentioned above are utilized as query models on resource-constrained end platforms.

\noindent \textbf{Feature Fusion}. Feature fusion has been widely studied in computer vision, \eg, detection~\cite{Zeng_2022_CVPR,Li_2022_CVPR}, multimedia retrieval~\cite{Shvetsova_2022_CVPR,LAFF}, \etc. As for image retrieval, it is broadly divided into three levels. The first is feature level, where features of different modalities~\cite{Croitoru_2021_ICCV}, scales~\cite{DOLG,DALG}, \etc, are effectively fused into a single feature. The second is indexing level, where multiple features are jointly indexed~\cite{zhang2013semantic,zheng2014packing}, or multiple visual vocabularies are fused together~\cite{zheng2014bayes}. The last is ranking level. Given several ranking lists returned by different retrieval algorithms, graph-based~\cite{zhang2012query}, or context-based~\cite{zheng2015query} methods fuse them into the final ranking list. 

However, all these methods require to extract multiple features on the query side. It inevitably increases the computational and storage complexity, which is hardly affordable for resource-constrained platforms. In contrast, we introduce a new asymmetric feature fusion paradigm, in which only a single lightweight model is deployed to embed queries and the gallery set is processed offline by various large powerful models on the cloud platforms. The proposed paradigm boosts the accuracy of asymmetric retrieval without adding any extra overhead to the query side.

\begin{figure*}[t]
	\begin{center}
		\resizebox{0.9\linewidth}{!}{
			\includegraphics[width=\linewidth]{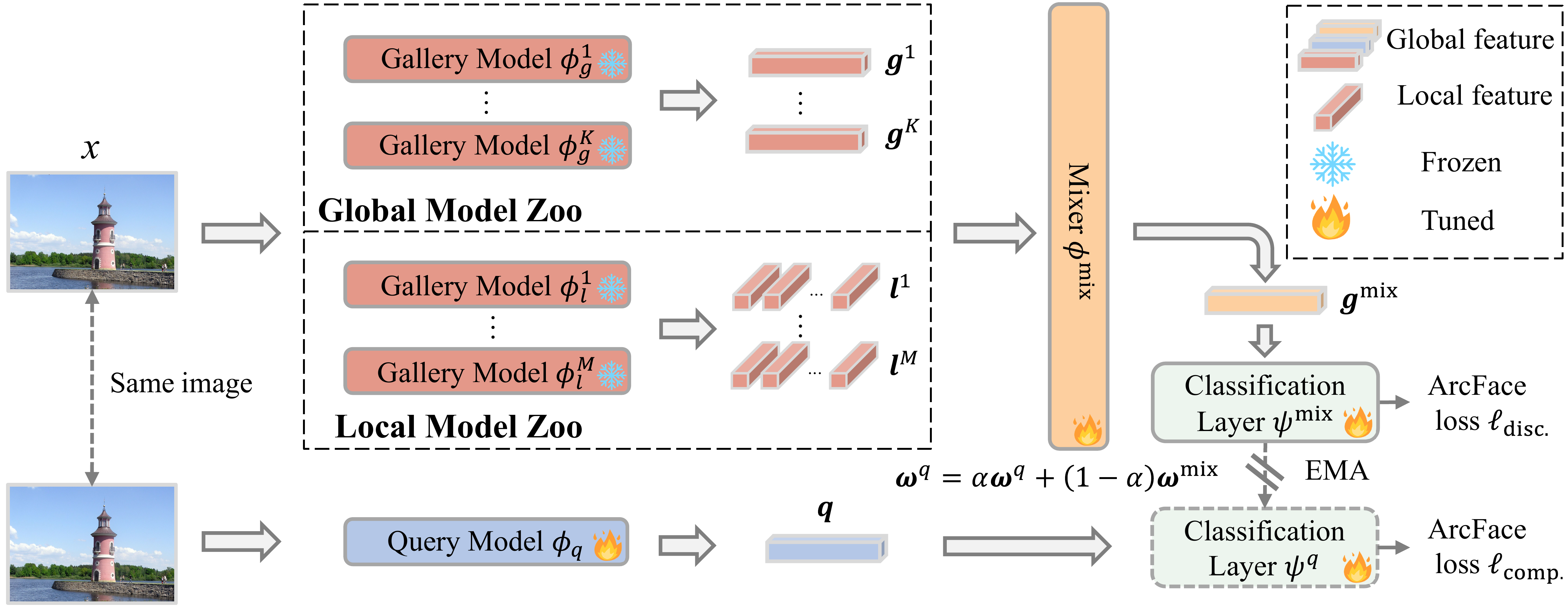}%
		}
	\end{center}
	\vspace*{-15pt}
	\caption{\textbf{Overview of the asymmetric feature fusion framework.}
		Given an image $x$, several models, \eg, global feature models $\{\phi_g^i: x \rightarrow \mathbb{R}^{D_i}\}_{i=1}^{K}$ and local feature models $\{\phi_l^i: x\rightarrow \mathbb{R}^{n_i \times d_i}\}_{i=1}^{M}$, are deployed on the gallery side to embed it into various features $\mathbf{G} = \{\bm{g}^i\in \mathbb{R}^{D_i}\}_{i=1}^K$ and $\mathbf{L} = \{\bm{l}^i\in \mathbb{R}^{n_i \times d_i}\}_{i=1}^{M}$. Then, a dynamic mixer (Sec.~\ref{sec:mixer}) is introduced to aggregate these features into a compact embedding $\bm{g}^{\rm mix} \in \mathbb{R}^{d}$, which is further fed into the classification layer $\psi^{\rm mix}$ for end-to-end optimization. On the query side, a lightweight model $\phi_q$ maps the same image $x$ to embedding $\bm{q} \in \mathbb{R}^{d}$. After that, $\bm{q}$ is fed into another classification layer $\psi^{q}$, a momentum-updated version of $\psi^{\rm mix}$, to train the query network for feature compatibility. Classification is adopted as the pretext task in the form of ArcFace~\cite{Deng_2019_CVPR} loss $\ell_{\rm disc.}$ and $\ell_{\rm comp.}$ (Sec.~\ref{sec:objective_function}) to train the mixer and the query network jointly.\vspace{-10pt}}
	\label{fig:training_overview}
\end{figure*}

\section{Preliminary on Asymmetric Retrieval}
\label{sec:background}

Asymmetric image retrieval aims to deploy models of different sizes on different platforms to realize search efficiency while preserving retrieval accuracy. Given a query set $\gQ$ and a gallery set $\gG$, query model $\phi_q: x\rightarrow\mathbb{R}^D$ and gallery model $\phi_g: x \rightarrow \mathbb{R}^D$ are deployed to embed them into $L_2$-normalized features, respectively. Then, the cosine similarities or Euclidean distances between query and gallery features are calculated to measure the similarities between images. Usually, an asymmetric retrieval system is expected to achieve similar accuracy as that of a symmetric retrieval system, \ie, $\gM(\phi_q(\gQ),\phi_g(\gG)) \approx \gM(\phi_g(\gQ),\phi_g(\gG))$,
where $\gM(\cdot,\cdot)$ denotes the evaluation metric for retrieval, \eg, mAP or Recall@K. 

Despite the promising performance achieved by existing asymmetric retrieval methods, we still observe notable retrieval accuracy degradation when compared to deploying large powerful models on both query and gallery sides (${\color{red}{\square}},{\color{orange}{\Diamond}},\dots,\color{blue}{\triangle}$ \vs $+$ in Fig.~\ref{fig:performace_compare}), \ie, $\gM(\phi_g(\gQ),\phi_g(\gG)) > \gM(\phi_q(\gQ),\phi_g(\gG))$.
This is due to the limited capacity of lightweight models, which cannot perfectly achieve feature compatibility with large powerful models. 

In this work, we alleviate the dilemma from the perspective of feature fusion and a new asymmetric feature fusion paradigm is introduced. Specifically, various large powerful models are deployed on the gallery side to extract features, which are further aggregated into compact embedding with a mixer. As for the query side, only a lightweight model is deployed, which is jointly trained with the mixer for feature compatibility. The proposed paradigm improves the accuracy of asymmetric retrieval systems without introducing any overhead to the resource-constrained query side.

\section{Asymmetric Feature Fusion}
\label{sec:method}
\subsection{Overview}
As shown in Fig.~\ref{fig:training_overview}, our AFF consists of multiple global feature models $\{\phi_g^i: x \rightarrow \mathbb{R}^{D_i}\}_{i=1}^{K}$ and local feature models $\{\phi_l^i: x\rightarrow \mathbb{R}^{N_i \times d_i}\}_{i=1}^{M}$ on the gallery side and a lightweight model $\phi_q: x \rightarrow \mathbb{R}^{d}$ on the query side. Let $\gT$ denote a training dataset. On the gallery side, each image $x$ in $\gT$ is first embedded into multiple global features $\mathbf{G} = \{\bm{g}^i\in \mathbb{R}^{D_i}\}_{i=1}^K$ and several sets of local features $\mathbf{L} = \{\bm{l}^i\in \mathbb{R}^{n_i \times d_i}\}_{i=1}^{M}$, respectively:
\begin{align}
	\bm{g}^i&=\phi_g^i(x) \in \mathbb{R}^{D_i},~i=1,2,\dots,K, \\
	\bm{l}^i&=\phi_l^i(x) \in \mathbb{R}^{n_i \times d_i},~i=1,2,\dots,M.
\end{align}

Typically, each local feature is associated with a coordinate tuple and a scale factor, indicating the location and image scale from which it is extracted. Our method ignores these information. All the global and local features are mapped to the same dimension of $d$ by the corresponding fully-connected layers:
\begin{align}
	\bm{f}_g^i&=\bm{g}^i\bm{W}_g^i \in \mathbb{R}^{d},~i=1,2,\dots,K, \\
	\bm{f}_l^i&=\bm{l}^i\bm{W}_l^i \in \mathbb{R}^{n_i \times d},~i=1,2,\dots,M.
\end{align}
After that, various gallery features are stacked together to form a feature sequence:
\begin{eqnarray}
	\label{equ:input_squence}
	\mathbf{F} = [\bm{f}_g^1;\dots;\bm{f}_g^K;\bm{f}_l^1;\dots;\bm{f}_l^M] \in \mathbb{R}^{N \times d},
\end{eqnarray}
where $N = K + \sum_{i=1}^{M}n_i$ is the total number of the gallery features. To reduce the storage overhead of the gallery side and improve search efficiency, a mixer $\phi_{\rm mix}:\mathbb{R}^{N \times d}\rightarrow \mathbb{R}^d$ (Sec.~\ref{sec:mixer}) is further introduced to transform $\mathbf{F}$ into compact embedding $\bm{g}^{\rm mix} = \phi_{\rm mix}(\mathbf{F}) \in \mathbb{R}^{d}$. On the query side, the same training image $x$ is embedded into $\bm{q}$ by the lightweight query model: $\bm{q} = \phi_{q}(x) \in \mathbb{R}^{d}$.

During training, the well-trained gallery models are kept frozen. Only the dynamic mixer $\phi_{\rm mix}$ and the query model $\phi_q$ are jointly trained for feature compatibility. The final objective function (Sec.~\ref{sec:objective_function}) consists of two losses:
\begin{eqnarray}
	\label{equ:objective_function}
	\mathop{\arg\min}_{\phi_{\rm mix}, \phi_q}~\frac{1}{| \gT |}\sum_{x\in {\gT}}(\ell_{\rm disc.}(\phi_{\rm mix},x) + \ell_{\rm comp.}(\phi_q, x)),
\end{eqnarray}
where $\ell_{\rm disc.}(\phi_{\rm mix};x)$ ensures the discrimination of the aggregated feature $\bm{g}^{\rm mix}$, and $\ell_{\rm comp.}(\phi_q;x)$ is designed to align query feature $\bm{q}$ and aggregated feature $\bm{g}^{\rm mix}$ in the same latent space so that they are mutually compatible.

\begin{figure}
	\centering
	
	\begin{subfigure}[b]{0.37\linewidth}
		\centering
		\begin{tikzpicture}[font={\footnotesize},]
	
	\node[] at (0, 125pt) (i1){$\mathbf{F}_{0,:}$};
	\node[] at (30pt, 125pt) (i2){$\cdots$};
	\node[] at (60pt, 125pt) (i3){$\mathbf{F}_{N-1,:}$};
	
	\node[op] at (30pt, 100pt) (cat){$+$};  
	
	\node[fill=blue!10,draw,minimum width=50pt,inner sep=2pt] at (30pt, 75pt) (d1){\Th{Linear}}; 
	\node[fill=purple!10,draw,minimum width=50pt,inner sep=2pt] at (30pt, 50pt) (gelu){\Th{GeLU}};
	\node[fill=blue!10,draw,minimum width=50pt,inner sep=2pt] at (30pt, 25pt) (d2){\Th{Linear}}; 
	\node[] at (30pt, 0pt) (out){$\bm{f}_{\text{fusion}}$}; 
	
	\draw[->] (i1) |- (cat);
	\draw[->] (i3) |- (cat);
	
	\node[dim] at (15pt,110pt) {$1 \times d$};
	\node[dim] at (75pt,110pt) {$1 \times d$};
	
	\draw[->] (cat) edge node[dim,midway,right]{$1 \times Nd$} (d1);
	\draw[->] (d1) edge node[dim,midway,right]{$1 \times Nd/2$} (gelu);
	\draw[->] (gelu) edge node[dim,midway,right]{$1 \times Nd/2$} (d2);
	\draw[->] (d2) edge node[dim,midway,right]{$1 \times d$} (out);
\end{tikzpicture}
		\vspace*{-12pt}
		\caption{Simple baseline}
		\label{fig:mixer_mlp}
	\end{subfigure}
	\hfill
	\begin{subfigure}[b]{0.57\linewidth}
		\centering
		\begin{tikzpicture}[font={\footnotesize},]
	
	\node[] at (0, 125pt) (fcls){$\bm{f}^{\rm fusion}$};  
	\node[] at (30pt, 125pt) (f2){$\mathbf{F}_{0,:}$}; 
	\node[] at (60pt, 125pt) (f3){$\cdots$}; 
	\node[] at (90pt, 125pt) (f4){$\mathbf{F}_{N-1,:}$};
	
	\coordinate(a1) at($(fcls.south) + (0, -1pt)$); 
	\coordinate(a2) at($(f2.south) + (0, -1pt)$); 
	\coordinate(a3) at($(f3.south) + (0, -3.0pt)$); 
	\coordinate(a4) at($(f4.south) + (0, -1pt)$); 
	
	\node[dim] at($(fcls.south) + (15pt, -6pt)$) {$1 \times d$};
	\node[dim] at($(f2.south) + (15pt, -6pt)$) {$1 \times d$};
	\node[dim] at($(f4.south) + (15pt, -6pt)$) {$1 \times d$};
	
	\node[op] at(45pt, 103pt) (cat){$+$};
	
	\draw[->] (a1) |- (cat);
	\draw[->] (a2) |- (cat);
	\draw[->] (a3) |- (cat);
	\draw[->] (a4) |- (cat);
	
	\node[self] at (45pt, 84pt) (self){MHSA};
	
	\coordinate(a5) at($(cat.south) + (0, -5pt)$);
	
	\coordinate(a6) at($(self.west) + (-10pt,0)$);
	
	\draw[->] (cat) edge (self);
	
	\node[addnorm] at (45pt, 69.0pt) (add){\Th{Add}~\&~\Th{Norm}};
	
	\draw[->] (self) edge (add);

	\node[fill=green!10,draw,minimum width=70pt,inner sep=2pt] at (45pt, 51pt) (ffn){FFN};
	
	\node[addnorm] at (45pt, 36pt) (add2){\Th{Add}~\&~\Th{Norm}};

	\draw (a5) -| (a6);
	\draw[->] (a6) |- (add);
	
	
	\draw[->] (add) edge (ffn);
	\draw[->] (ffn) edge (add2);
	
	\coordinate(a12) at($(add.south) + (0, -4.0pt)$);
	
	\coordinate(a13) at($(ffn.west) + (-10pt,0)$);
	
	\draw (a12) -| (a13);
	\draw[->] (a13) |- (add2);
	
	\coordinate(a7) at($(add2.south) + (0pt,-10pt)$);
	
	\node[] at (0, 0pt) (ocls){$\widetilde{\bm{f}}_{\rm fusion}$};  
	\node[] at (30pt, 0pt) (o2){$\widetilde{\mathbf{F}}_{0,:}$}; 
	\node[] at (60pt, 0pt) (o3){$\cdots$}; 
	\node[] at (90pt, 0pt) (o4){$\widetilde{\mathbf{F}}_{N-1,:}$};
	
	\coordinate(a8) at($(ocls.north) + (0, 0.2pt)$); 
	\coordinate(a9) at($(o2.north) + (0, 0.2pt)$); 
	\coordinate(a10) at($(o3.north) + (0, 3.8pt)$); 
	\coordinate(a11) at($(o4.north) + (0, 0.2pt)$); 

	\draw[-] (add2) edge (a7);
	\draw[->] (a7) -| (a8);
	\draw[->] (a7) -| (a9);
	\draw[->] (a7) -| (a10);
	\draw[->] (a7) -| (a11);
	
	\node[dim] at($(ocls.south) + (15pt, 21pt)$) {$1 \times d$};
	\node[dim] at($(o2.south) + (15pt, 21pt)$) {$1 \times d$};
	\node[dim] at($(o4.south) + (15pt, 21pt)$) {$1 \times d$};
	
	\node[draw=black!70, minimum width=108pt,minimum height=70pt,dashed,fill opacity=.3,text opacity=1, line width=0.6pt] at (45pt, 62pt) (outer){};
	\node[black!60] at (90pt, 32pt) (case){$\times C$}; 
	
\end{tikzpicture}
		\vspace*{-12pt}
		\caption{Ours}
		\label{fig:mixer_tr}
	\end{subfigure}
	\vspace*{-5pt}
	\caption{\textbf{Different variants of mixer}. (a) Simple baseline: different features are concatenated, followed by dimension reduction with several fully-connected layers. (b) Our mixer: a fusion token and the feature sequence are iteratively processed by a transformer layer~\cite{Transformer}, where the fusion token dynamically aggregates beneficial features from various gallery features.\vspace*{-5pt}}
	\label{fig:mixer}
	
\end{figure}
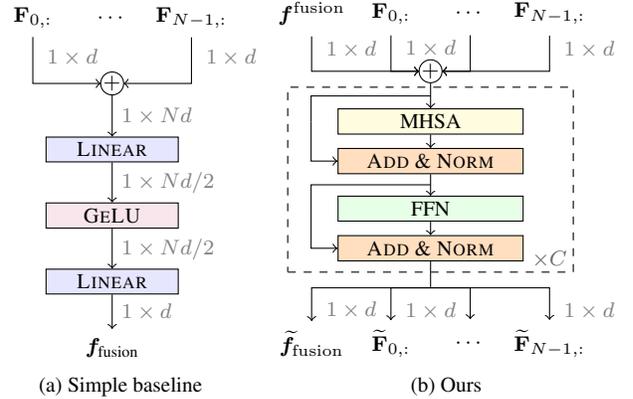

\subsection{Dynamic Mixer} \label{sec:mixer} 
Given an image encoded by various features $\mathbf{F}$, feature fusion aims to combine those features for better retrieval accuracy. A simple way is to concatenate various features and perform dimension reduction, which is implemented by several fully-connected layers (Fig.~\ref{fig:mixer_mlp}). However, it leads to an over-parameterized mixer when the number of features is large, which may cause overfitting. 

In this work, attention mechanism~\cite{Transformer} is adopted to aggregate various features (Fig.~\ref{fig:mixer_tr}). A learnable fusion token $\bm{f}_{\rm fusion} \in \mathbb{R}^d $ is first added to the top of $\mathbf{F}$ to form the input:
\begin{align}
	\mathbf{F}_{\rm input} = [\bm{f}_{\rm fusion};\mathbf{F}] \in \mathbb{R}^{(N+1) \times d}.
\end{align}
Then, $\mathbf{F}_{\rm input}$ is iteratively processed $C$ times by a  transformer layer, which is formulated as:
\begin{equation}\label{equ:transformer}
	\begin{aligned}
		\mathbf{\bar{Z}}^{i+1} &=\textsc{LN}(\mathbf{Z}^i + \textsc{MHSA}(\mathbf{Z}^i)),\\
		\mathbf{Z}^{i+1} &= \textsc{LN}(\mathbf{\bar{Z}}^{i+1} + \textsc{MLP}(\mathbf{\bar{Z}}^{i+1})),\\
		\textsc{MLP}(\mathbf{\bar{Z}}^{i+1}) &= \textsc{GeLU}(\mathbf{\bar{Z}}^{i+1}\bm{W}_1)\bm{W}_2,\\
		i &= 0, \dots, C-1,
	\end{aligned}
\end{equation}
where $\mathbf{Z}^0 = \mathbf{F}_{\rm input}$; MHSA is the Multi-Head Self-Attention~\cite{Transformer}; $\textsc{MLP}$ is a two-layer perceptron with parameter matrices $\bm{W}_1 \in \mathbb{R}^{d \times d_e}$ and $\bm{W}_2 \in \mathbb{R}^{d_e \times d}$, and an intermediate dimension $d_e = 2 \times d$; $\textsc{LN}$ is the layer normalization~\cite{ba2016layer}. The final output fusion token $\widetilde{\bm{f}}_{\rm fusion} = \mathbf{Z}^{C-1}_{0,:}$ is adopted as the aggregated feature $\bm{g}^{\rm mix}$ for the gallery side.

\subsection{Training Objective Functions} \label{sec:objective_function} 

To ensure the superiority of our asymmetric feature fusion paradigm, there are two requirements needed to be guaranteed. First, the aggregated feature $\bm{g}^{\rm mix}$ is expected to be more discriminative than any single gallery feature. To this end, following the state-of-the-art metric learning methods~\cite{DELG,DOLG,CVNet} in image retrieval, classification is adopted as a pretext task in the form of ArcFace loss~\cite{Deng_2019_CVPR} to train the mixer. Assuming the classification layer $\psi^{\rm mix}: \mathbb{R}^d \rightarrow \mathbb{R}^N$, with $N$ categories, is parameterized by weights $\bm{\omega}^{\rm mix} \in \mathbb{R}^{N \times d}$, the loss is formulated as: 
\begin{equation}\label{eq:mixer_arcface_loss}
	\begin{split}
		\! \ell_{\rm disc.}(\phi_{\rm mix}, x) \!= \!- \log{ \frac{e^{s \cdot {\rm cos}(\theta_y^1 + m)}}{e^{s \cdot {\rm cos}(\theta_y^1 + m)} + \sum\limits_{j\neq y}e^{s \cdot {\rm cos}(\theta_j^1)} }}, 
	\end{split}
\end{equation}
where $y$ is the label of the training image $x$, $s$ is a scale factor, $m$ is the margin, and $\theta_y^1 = {\rm arccos}(\langle \frac{\bm{\omega}^{\rm mix}_{y,:}}{\|\bm{\omega}^{\rm mix}_{y,:}\|}, \bm{g}^{\rm mix}\rangle)$ is the angle between the $y$-th $L_2$-normalized prototype of the classifier $\psi^{\rm mix}$ and the feature $\bm{g}^{\rm mix}$.

Second, query feature $\bm{q}$ and aggregated feature $\bm{g}^{\rm mix}$ should be compatible with each other. One may share the same classifier between $\phi^{\rm mix}$ and $\phi_q$, which has shown effectiveness in previous methods~\cite{BCT,HVS,LCE}. However, our approach expects to train $\phi^{\rm mix}$ and $\phi_q$ jointly. Simply sharing classifier couples the training of networks with different capabilities, which may damage the discriminative capability of the aggregated embedding $\bm{g}^{\rm mix}$. Besides, the classifier parameters $\bm{\omega}^{\rm mix}$ evolve rapidly, which cannot provide a stable target to the query model. To this end, we decouple the training processes of $\phi_{\rm mix}$ and  $\phi_q$, while ensuring feature compatibility through a momentum update mechanism. ArcFace loss is still adopted for training query model $\phi_q$:
\begin{equation}\label{eq:query_net_arcface_loss}
	\begin{split}
		\ell_{\rm comp.}(\phi_q, x) = - \log{ \frac{e^{s \cdot {\rm cos}(\theta_y^2 + m)}}{e^{s \cdot {\rm cos}(\theta_y^2 + m)} + \sum\limits_{j\neq y}e^{s \cdot {\rm cos}(\theta_j^2)} }}, 
	\end{split}
\end{equation}
where $\theta_y^2 = {\rm arccos}(\langle \frac{\bm{\omega}^{q}_{y,:}}{\|\bm{\omega}^{q}_{y,:}\|}, \bm{q} \rangle)$ is the angle between the $y$-th $L_2$-normalized prototype of the classifier $\psi^{q}$ and the feature $\bm{q}$.
Differently, the parameter $\bm{\omega}^q$ is not updated through back-propagation, but a moving-averaged version of $\bm{\omega}^{\rm mix}$:
\begin{eqnarray}\label{equ:momentum_update}
	\bm{\omega}^q \leftarrow \alpha \bm{\omega}^q + (1 - \alpha) \bm{\omega}^{\rm mix},
\end{eqnarray}
where $\alpha \in [0,1)$ is a momentum coefficient. Only the parameters $\bm{\omega}^{\rm mix}$ are updated by back-propagation. This momentum update in Eq.~(\ref{equ:momentum_update}) decouples the training of $\phi_{\rm mix}$ and $\phi_q$ while making $\bm{\omega}^q$ evolve more smoothly than $\bm{\omega}^{\rm mix}$.


\section{Experiments}
\label{sec:experiments}
\subsection{Experimental Setup}

\mypartight{Evaluation Datasets and Metrics}. We evaluate the proposed framework on three landmark retrieval datasets, including GLDv2-Test~\cite{GLDv2}, Revisited Oxford ($\gR$Oxf), Revisited Paris ($\gR$Par)~\cite{roxpar}. GLDv2-Test contains $761,757$ gallery images, and $390$/$750$ images as public/private query sets, respectively. The evaluation metric is mAP$@100$. As for $\gR$Oxf and $\gR$Par, there are $70$ queries for both of them, with $4,993$ and $5,007$ gallery images, respectively. mAP on the Medium and Hard settings are reported. Large-scale results are reported with the $\gR$1M~\cite{roxpar} dataset added (+ 1M).

\mypartight{Gallery and Query models}. Four global features including DELG~\cite{DELG}, Token~\cite{Token}, DOLG~\cite{DOLG}, CVNet~\cite{CVNet} and two local features HOW~\cite{HOW}, $^\star$DELG~\cite{DELG} are adopted as gallery features. As for query model, we only keep the feature extractor of lightweight models, \eg, ShuffleNets~\cite{shufflenetv2} and MobileNets~\cite{Mobilenetv2}, with a GeM pooling~\cite{gem} layer and a whitening layer added at the end.

\mypartight{Training Details}. GLDv2~\cite{GLDv2} is adopted for training, which consists of $1,580,470$ images with $81,311$ classes. All the gallery features are extracted offline for training efficiency. During training, a $512 \times 512$-pixel region is cropped from each randomly resized training image, followed by random color jittering. We jointly train the mixer and query model for 20 epochs on four NVIDIA RTX 3090 GPUs with a batch size of $128$. SGD is adopted as the optimizer with a weight decay of $0.01$ and an initial learning rate of $0.001$, which linearly decays to $0$ when the desired number of steps is reached. $C$ in Eq.~(\ref{equ:transformer}) is set to $4$. $d$ and $d_e$ in Eq.~(\ref{equ:transformer}) are both set to $2,048$. Margin $m$ and scale $s$ in Eq.~(\ref{eq:mixer_arcface_loss}) and Eq.~(\ref{eq:query_net_arcface_loss}) are set as $0.3$ and $32.0$, respectively.

\subsection{Ablation Study}
\label{sub:ablations} 
\noindent \textbf{Variants of the mixer.} In Tab.~\ref{tab:mixer_type}, we compare our proposed mixer against simple baseline (MLP), LAFF~\cite{LAFF}, and OrthF.~\cite{DOLG}. Our transformer-based mixer iteratively extracts useful information from various gallery features via a fusion token, which achieves the best retrieval accuracy.

\begin{table}[tb]
	\centering
	\resizebox{1\linewidth}{!}
	{
		\small
		\setlength{\extrarowheight}{-1.5pt}
		\setlength{\tabcolsep}{1.5pt}
		\begin{tabular}{lc cc cc cc}
			\toprule
			\mc{1}{\mr{2}{ \tabincell{c}{ \Th{Mixer} \\ \Th{Type}} }} & \mr{2}{ \tabincell{c}{ \Th{Query} \\ \Th{Net} $\phi_q$} } &\mc{2}{GLDv2-Test} & \multicolumn{2}{c}{$\gR$Oxf + 1M} & \multicolumn{2}{c}{$\gR$Par + 1M} \\
			
			\cmidrule(l){3-4}  \cmidrule(l){5-6}  \cmidrule(l){7-8}
			
			& & Private & Public & Medium & Hard & Medium & Hard \\
			
			\midrule
			
			~~\mr{2}{MLP}                           & Mixer & 31.13 & 29.63 & 74.89 & 53.79 & 79.97 & 62.25 \\
			& MobV2 & 27.63 & 24.36 & 65.61 & 42.84 & 71.09 & 52.57 \\
			\cdashlinelr{1-8} 
			~~\mr{2}{LAFF}              & Mixer & 31.90 & 30.07 & 76.30 & 55.68 & 81.38 & 64.31 \\
			& MobV2 & 28.43 & 26.48 & 66.53 & 45.10 & 73.96 & 54.37 \\
			\cdashlinelr{1-8} 
			~~\mr{2}{OrthF.}            & Mixer & 30.27 & 29.33 & 73.78 & 52.00 & 79.34 & 61.84 \\
			& MobV2 & 26.73 & 24.52 & 63.38 & 38.86 & 70.70 & 51.26 \\
			\cdashlinelr{1-8} 
			~~\mr{2}{\textbf{Ours}} & Mixer & \bf{32.85} & \bf{31.27} & \bf{77.84} & \bf{58.91} & \bf{84.43} & \bf{69.44} \\
			& MobV2 & \bf{\color{Maroon!80}{29.85}} & \bf{\color{Maroon!80}{27.68}} & \bf{\color{Maroon!80}{70.47}} & \bf{\color{Maroon!80}{49.16}} & \bf{\color{Maroon!80}{80.01}} & \bf{\color{Maroon!80}{62.58}} \\
			\bottomrule
		\end{tabular} 
	}
	\vspace*{-0.5em}
	\caption{Analysis of \textbf{different mixer architectures}. Mixer: query images are embedded into various features, which are further aggregated into a compact vector; MobV2: MobileNetV2 is deployed to embed query images. Mixer is also adopted as gallery model. LAFF:~\cite{LAFF}; OrthF.:~\cite{DOLG}.\vspace*{-10pt}}\label{tab:mixer_type}
\end{table}

\noindent \textbf{Impact of different gallery features}. In Tab.~\ref{tab:combinations}, we investigate how our method responds when different gallery features are added. Results show that each feature is beneficial for the task and features have different effects when retrieval is performed in different gallery sets. Our method dynamically aggregates diverse features to enhance the representation of gallery images.

\begin{table}[tb]
	\centering
	\resizebox{1\linewidth}{!}
	{
		\small
		\setlength{\extrarowheight}{-1.5pt}
		\setlength{\tabcolsep}{1.0pt}
		\begin{tabular}{cccc c cc cc cc}
			\toprule
			\mr{2}{\rotatebox{90}{DELG}} & \mr{2}{\rotatebox{90}{Token}} & \mr{2}{\rotatebox{90}{DOLG}} & \mr{2}{\rotatebox{90}{CVNet}} & \mr{2}{ \tabincell{c}{ \Th{Query} \\ \Th{Net} $\phi_q$} }  & \mc{2}{GLDv2-Test} & \mc{2}{$\gR$Oxf + 1M} & \mc{2}{$\gR$Par + 1M} \\
			
			\cmidrule(l){6-7}  \cmidrule(l){8-9}  \cmidrule(l){10-11}
			
			& & & & & Private & Public & Medium & Hard & Medium & Hard \\
			
			\midrule
			
			\mr{2}{\checkmark} &                    & \mr{2}{\checkmark} & \mr{2}{\checkmark} & Mixer & 32.31 & 29.77 & 75.49 & 54.96 & 83.05 & 67.21 \\
			&                    &                    &                    & MobV2 & 28.53 & 25.39 & 66.49 & 44.83 & 77.19 & 59.08 \\
			\cdashlinelr{1-11}
			
			\mr{2}{\checkmark} & \mr{2}{\checkmark} &                    & \mr{2}{\checkmark} & Mixer & 32.09 & 30.31 & 76.28 & 55.86 & 80.97 & 62.99 \\
			&                    &                    &                    & MobV2 & 27.45 & 25.75 & 66.08 & 44.51 & 73.18 & 53.73 \\
			\cdashlinelr{1-11}
			\mr{2}{\checkmark} & \mr{2}{\checkmark} & \mr{2}{\checkmark} &                    & Mixer & 31.83 & 30.07 & 76.30 & 55.68 & 81.38 & 64.31 \\
			&                    &                    &                    & MobV2 & 26.31 & 25.34 & 66.10 & 44.76 & 75.11 & 55.02 \\
			\cdashlinelr{1-11}
			& \mr{2}{\checkmark} & \mr{2}{\checkmark} & \mr{2}{\checkmark} & Mixer & 32.82 & 30.99 & 77.71 & \bf{59.28} & 82.86 & 66.95 \\
			&                    &                    &                    & MobV2 & 28.35 & 26.40 & 70.20 & \bf{\color{Maroon!80}{49.29}} & 76.91 & 57.47 \\
			\cdashlinelr{1-11}
			\mr{2}{\checkmark} & \mr{2}{\checkmark} & \mr{2}{\checkmark} & \mr{2}{\checkmark} & Mixer & \bf{32.85} & \bf{31.27} & \bf{77.84} & 58.91 & \bf{84.43} & \bf{69.44} \\
			&                    &                    &                    & MobV2 & \bf{\color{Maroon!80}{29.85}} & \bf{\color{Maroon!80}{27.68}} & \bf{\color{Maroon!80}{70.47}} & 49.16 & \bf{\color{Maroon!80}{80.01}} & \bf{\color{Maroon!80}{62.58}} \\
			\bottomrule
		\end{tabular} 
	}
	\vspace*{-0.5em}
	\caption{Analysis of \textbf{different feature combinations}. Mixer: query images are embedded into various features, which are further aggregated into a compact vector; MobV2: MobileNetV2 is deployed to embed query images. Mixer is adopted as gallery model under all settings.\vspace*{-15pt}}\label{tab:combinations}
\end{table}

\noindent \textbf{Robustness}. When it comes to feature fusion, there is usually no idea whether a specific feature is beneficial for the current task or not in real-world applications. Feature fusion inevitably introduces some noisy features. 

In Tab.~\ref{tab:robustness}, we investigate the robustness of the proposed framework to noisy features. First, three feature extractors with unsatisfactory performance including RGeM~\cite{gem}, VGeM~\cite{gem}, and RMAC~\cite{RMAC}, are added to the gallery model zoo. These noisy features cause significant performance degradation for direct feature ensemble, \eg, mAP on $\gR$Oxf + 1M drops from $75.30$ to $71.52$. Our method is almost unaffected. We further consider an extreme case, in which \textbf{Gaussian White Noise} is directly added to the gallery feature set. This further leads to a huge performance degradation for feature ensemble, \ie, mAP on $\gR$Oxf + 1M drops from $71.52$ to $57.78$. Our method also suffers a slight performance degradation in this setup. To summarize, our mixer dynamically extracts informative features from diverse gallery features by a fusion token, which demonstrates superior robustness to noise.
 
\begin{table}[tb]
	\centering
	\resizebox{1\linewidth}{!}
	{
		\scriptsize
		\setlength{\extrarowheight}{-1.5pt}
		\setlength{\tabcolsep}{1.0pt}
		\begin{tabular}{c ccccc cc cc cc}
			\toprule 
			\mr{2}{\rotatebox{90}{\emph{\color{gray}{Default}}}} & \mr{2}{\rotatebox{90}{RGeM}} & \mr{2}{\rotatebox{90}{VGeM}} & \mr{2}{\rotatebox{90}{RMAC}} & \mr{2}{\rotatebox{90}{Noisy}} & \mr{2}{\Th{Method}} & \mc{2}{GLDv2-Test} & \multicolumn{2}{c}{$\gR$Oxf + 1M} & \multicolumn{2}{c}{$\gR$Par + 1M} \\
			
			\cmidrule(lr){7-8} \cmidrule(lr){9-10}  \cmidrule(lr){11-12}
			
			& & & & & & Private & Public & Medium & Hard & Medium & Hard \\
			
			\midrule
			\mr{2}{\checkmark} &  &  &  & & Ensemble & 32.28 & 29.76 & 75.30 & 53.24 & 83.35 & 67.05 \\
			&                    &                    &                    &                    & Mixer    & \bf{32.85} & \bf{31.27} & \bf{77.84} & \bf{58.91} & \bf{84.43} & \bf{69.44} \\
			\cdashlinelr{1-12}
			& \mr{2}{\checkmark} & \mr{2}{\checkmark} & \mr{2}{\checkmark} & & Ensemble & 17.18 & 15.71 & 51.95 & 26.65 & 56.19 & 28.69 \\
			&                    &                    &                    &                    & Mixer    & \bf{20.84} & \bf{19.11} &\bf{57.85} & \bf{29.55} & \bf{62.92} & \bf{37.54} \\
			\cdashlinelr{1-12}
			\mr{2}{\checkmark} & \mr{2}{\checkmark} & \mr{2}{\checkmark} & \mr{2}{\checkmark} & & Ensemble & 29.29 & 26.90 & 71.52 & 45.27 & 76.74 & 55.77 \\
			&                    &                    &                    &                    & Mixer    & \bf{32.87} & \bf{31.00} & \bf{77.28} & \bf{58.03} & \bf{83.56} & \bf{68.05} \\
			\cdashlinelr{1-12}
			\mr{2}{\checkmark} &                    &                    &                    & \mr{2}{\checkmark} & Ensemble & 29.32 & 26.88 & 52.30 & 25.67 & 54.44 & 26.52 \\
			&                    &                    &                    &                    & Mixer    & \bf{32.60} & \bf{30.94} & \bf{76.40} & \bf{57.13} & \bf{82.65} & \bf{66.52} \\
			\cdashlinelr{1-12}
			\mr{2}{\checkmark} & \mr{2}{\checkmark} & \mr{2}{\checkmark} & \mr{2}{\checkmark} & \mr{2}{\checkmark} & Ensemble & 30.24 & 28.47 & 57.78 & 29.88 & 59.88 & 32.58 \\
			&                    &                    &                    & & Mixer    & \bf{32.56} & \bf{30.61} & \bf{77.65} & \bf{56.96} & \bf{82.68} & \bf{66.49} \\
			\bottomrule
		\end{tabular} 
	}
	\vspace*{-5pt}
	\caption{ \textbf{Robustness analysis (symmetric retrieval)}. \emph{\color{gray}{Default}}: four global features including DELG~\cite{DELG}, Token~\cite{Token}, DOLG~\cite{DOLG} and CVNet~\cite{CVNet}. RGeM~\cite{gem}, VGeM~\cite{gem} and RMAC~\cite{gem} are three global features with weak retrieval accuracy. Noisy is randomly sampled from a i.i.d $D$-dimensional Gaussian distribution. Ensemble: feature concatenation; Mixer: various features are aggregated into a compact vector with our proposed mixer.\vspace*{-10pt}}\label{tab:robustness}
\end{table}

\noindent \textbf{Larger gallery models}. Considering that feature fusion leads to a significant increase in the number of parameters in the gallery model. This raises the question of whether similar performance can be achieved by using a larger model on the gallery side instead. As shown in Tab.~\ref{tab:large_backbone}, simply using a larger model does not necessarily result in better retrieval accuracy. Moreover, larger models typically entail a significant training overhead. In contrast, our approach only requires training a lightweight fusion network while keeping the existing feature extraction network frozen.

\begin{table}[tb]
	\centering
	\resizebox{1.0\linewidth}{!}
	{
		\footnotesize
		\setlength{\extrarowheight}{-1.0pt}
		\setlength{\tabcolsep}{0.5pt}
		\begin{tabular}{cccc cc cc cc}
			\toprule
			\mr{2}{\Th{Method}}&\mr{2}{\tabincell{c}{ \Th{Query} \\ ~~~\Th{Net} $\phi_q$}} & \mr{2}{\tabincell{c}{ \Th{Gallery} \\ ~~\Th{Net} $\phi_g$}}   & \mc{2}{GLDv2-Test} & \mc{2}{$\gR$Oxf + 1M} & \mc{2}{$\gR$Par + 1M} \\
			
			\cmidrule(l){4-5}  \cmidrule(l){6-7}  \cmidrule(l){8-9}
			
			& & & Private & Public & Medium & Hard & Medium & Hard \\
			
			\midrule
			
			\mr{2}{~\tabincell{c}{ DOLG~\cite{DOLG} \\ +CSD~\cite{CSD}}} & Swin-B & Swin-B & 29.69 & 27.63 & 75.31 & 53.67 & 81.19 & 64.29 \\
			& MobV2  & Swin-B & \bf{\color{gray}{26.98}} & \bf{\color{gray}{24.34}} & \bf{\color{gray}{66.21}} & \bf{\color{gray}{43.34}} & \bf{\color{gray}{70.76}} & \bf{\color{gray}{50.51}} \\
			
			\cdashlinelr{1-9}
			
			\mr{2}{~\tabincell{c}{ DOLG~\cite{DOLG} \\ +CSD~\cite{CSD}}} & Swin-L & Swin-L & 28.42 & 26.27 & 72.72 & 50.55 & 81.43 & 62.52 \\
			& MobV2  & Swin-L & 25.65 & 23.34 & 64.26 & 39.82 & 69.79 & 46.53 \\
			\cdashlinelr{1-9} 
			
			\textbf{Ours} & Mixer  & Mixer  & \bf{32.93} & \bf{31.63} & \bf{77.58} & \bf{58.30} & \bf{83.68} & \bf{68.04} \\
			\textbf{Ours} & MobV2  & Mixer  & \bf{\color{Maroon!80}{28.57}} & \bf{\color{Maroon!80}{26.57}} & \bf{\color{Maroon!80}{68.17}} & \bf{\color{Maroon!80}{47.27}} & \bf{\color{Maroon!80}{77.90}} & \bf{\color{Maroon!80}{59.01}} \\
			\bottomrule
		\end{tabular} 
	}
	\vspace*{-5pt}
	\caption{\textbf{Comparison of different gallery models}. Feature dimension is set as $512$. Swin-B (88 MB) and Swin-L (197 MB): base and large version of Swin Transformer~\cite{Liu_2021_ICCV}; Mixer (202 MB): DOLG~\cite{DOLG}, Token~\cite{Token}, CVNet~\cite{CVNet} and DELG~\cite{DELG} are adopted for feature fusion.\vspace*{-10pt}}
	\label{tab:large_backbone}
\end{table}

\noindent \textbf{Broad applicability}. A common image retrieval practice usually first retrieves candidates via similarity search with global features and then re-ranks with corresponding local features. However, the overhead of extracting local features is much higher than that of global features, which is unaffordable for resource-constrained platforms. Here, we explore the effectiveness of fusing global and local features only on the gallery side.

In Tab.~\ref{tab:local_agg}, our method is compared with the traditional ASMK~\cite{ASMK}-based local feature aggregation approaches~\cite{HOW,Superfeature}. Even under the asymmetric setting, it achieves better accuracy with less overhead for the query side. In Tab.~\ref{tab:global_local_agg}, our method also achieves higher accuracy than re-ranking~\cite{oxford,RRT} methods under the asymmetric setting. Note that the re-ranking process introduces significant retrieval latency, while the proposed paradigm only needs efficient vector search. To summarize, our method aggregates diverse features on the gallery side, which greatly reduces the overhead for the query side and online retrieval latency while improving retrieval accuracy.

\begin{table}[tb]	
	\begin{subtable}{1\linewidth}
		\centering
		\scriptsize
		\setlength{\extrarowheight}{0.0pt}
		\setlength{\tabcolsep}{0.6pt}
		\resizebox{1\linewidth}{!}
		{
			\begin{tabular}{ccc c cc cc cc cc}
				\toprule 
				\mr{2}{\rotatebox{90}{DELG}} & \mr{2}{\rotatebox{90}{$^\star$DELG}} & \mr{2}{\rotatebox{90}{HOW}} & \mr{2}{ \Th{Method} } & \Th{Ret}. & ~\Th{Ext}. &  \mc{2}{GLDv2-Test} & \multicolumn{2}{c}{$\gR$Oxf + 1M} & \multicolumn{2}{c}{$\gR$Par + 1M} \\
				
				\cmidrule(lr){5-6} \cmidrule(lr){7-8}  \cmidrule(lr){9-10} \cmidrule(lr){11-12}
				
				& & & & (\emph{s}) $\downarrow$ & (\emph{ms}) $\downarrow$ & Private & Public & Medium & Hard & Medium & Hard \\
				
				\midrule
				& \checkmark & & ASMK$^*$ & 1.042 & ~388.9 & 20.21 & 17.47 & 62.78 & 38.59 & 66.71 & 40.89 \\
				& \checkmark & & \textbf{Ours} & 0.345  & ~404.1 & \bf{26.76} & \bf{24.34} & \bf{66.53} & \bf{42.97} & \bf{73.02} & \bf{51.21} \\
				& \checkmark & & ~Ours$^{\ddagger}$ &  \bf{0.345} & ~\bf{16.5} & \bf{\color{Maroon!80}{23.66}} & \bf{\color{Maroon!80}{21.66}} & \bf{\color{Maroon!80}{60.94}} & \bf{\color{Maroon!80}{41.35}} & \bf{\color{Maroon!80}{69.89}} & \bf{\color{Maroon!80}{46.85}} \\
				\cdashlinelr{1-12}
				& & \checkmark & ASMK$^*$   & 0.995 &  ~258.1  & 16.52 & 14.05 & 63.66 & 36.84 & 58.42 & 30.73 \\
				& & \checkmark & \textbf{Ours} & 0.345  & ~273.2 & \textbf{23.22} & \textbf{22.32} & \bf{68.60} & \bf{41.77} & \bf{67.08} & \bf{46.96} \\
				& & \checkmark & ~Ours$^{\ddagger}$ &  \bf{0.345} &   ~\bf{16.5} & \bf{\color{Maroon!80}{22.08}} & \bf{\color{Maroon!80}{20.16}} & \bf{\color{Maroon!80}{62.36}} & \bf{\color{Maroon!80}{40.07}} & \bf{\color{Maroon!80}{64.92}} & \bf{\color{Maroon!80}{41.64}} \\
				\cdashlinelr{1-12}
				& \checkmark & \checkmark & ASMK$^*$     & 2.324  & ~647.2 & 20.35 & 17.28 & 68.92 & 40.70 & 68.06 & 42.43 \\
				& \checkmark & \checkmark & \textbf{Ours}       & 0.345  & ~665.1 & \bf{27.24} & \bf{25.72} & \bf{72.40} & \bf{49.86} & \bf{76.78} & \bf{57.03} \\
				& \checkmark & \checkmark & ~Ours$^{\ddagger}$      & \bf{0.345} & ~\bf{16.5} & \bf{\color{Maroon!80}{25.06}} & \bf{\color{Maroon!80}{23.28}} & \bf{\color{Maroon!80}{65.18}} & \bf{\color{Maroon!80}{44.23}} & \bf{\color{Maroon!80}{71.49}} & \bf{\color{Maroon!80}{48.76}} \\
				
				\cdashlinelr{1-12}
				\checkmark & \checkmark & \checkmark & \textbf{Ours}       &  0.345 &  ~678.7    & \bf{28.13} & \bf{26.95} & \bf{73.71} & \bf{51.00} & \bf{78.60} & \bf{59.25} \\
				\checkmark & \checkmark & \checkmark & ~Ours$^{\ddagger}$ & \bf{0.345} &  ~\bf{16.5} & \bf{\color{Maroon!80}{26.38}} & \bf{\color{Maroon!80}{24.22}} & \bf{\color{Maroon!80}{66.48}} & \bf{\color{Maroon!80}{46.96}} & \bf{\color{Maroon!80}{73.55}} & \bf{\color{Maroon!80}{51.89}} \\
				
				\bottomrule
			\end{tabular} 
		}
		\caption{Aggregating local features into compact embeddings.}%
		\label{tab:local_agg}
	\end{subtable}
	\vskip 3pt
	\begin{subtable}{1\linewidth}
		\centering
		\scriptsize
		\setlength{\extrarowheight}{0.0pt}
		\setlength{\tabcolsep}{0.6pt}
		\resizebox{1\linewidth}{!}
		{
			\begin{tabular}{ccc c cc cc cc cc}
				\toprule 
				\mr{2}{\rotatebox{90}{DELG}} & \mr{2}{\rotatebox{90}{$^\star$DELG}} & \mr{2}{\rotatebox{90}{HOW}} & \mr{2}{ \Th{Method} } & \Th{Ret}. & ~\Th{Ext}. &  \mc{2}{GLDv2-Test} & \multicolumn{2}{c}{$\gR$Oxf + 1M} & \multicolumn{2}{c}{$\gR$Par + 1M} \\
				
				\cmidrule(lr){5-6} \cmidrule(lr){7-8}  \cmidrule(lr){9-10} \cmidrule(lr){11-12}
				
				& & & & (\emph{s}) $\downarrow$ & (\emph{ms}) $\downarrow$ & Private & Public & Medium & Hard & Medium & Hard \\
				
				\midrule
				\checkmark & & &  GR            & 0.345  & 113.7  & 26.14 & 24.20 & 63.71 & 37.45 & 70.59 & 46.94 \\
				\checkmark & & & ~GR$^{\ddagger}$                      & \bf{0.345} & \bf{16.5}   & 23.33 & 21.33 & 60.19 & 34.14 & 68.25 & 44.24 \\
				\cdashlinelr{1-12}
				\checkmark & \checkmark & & GR + GV$^\dagger$  & 12.752  & 425.4 & 26.81 & 24.73 & 68.63 & 45.33 & 71.38 & 47.78 \\
				\checkmark & \checkmark & & GR + RRT$^\dagger$ &  1.875  & 425.4 & 27.94 & 26.05 & 69.01 & 45.47 & 72.64 & 49.93 \\
				\checkmark & \checkmark & & \textbf{Ours}      & 0.345  &  441.3  & \bf{28.71} & \bf{26.79} & \bf{69.28} & \bf{46.82} & \bf{74.76} & \bf{54.15} \\
				\checkmark & \checkmark & & ~Ours$^{\ddagger}$               & \bf{0.345} & \bf{16.5} & \bf{\color{Maroon!80}{26.25}} & \bf{\color{Maroon!80}{24.18}} & \bf{\color{Maroon!80}{62.00}} & \bf{\color{Maroon!80}{40.31}} & \bf{\color{Maroon!80}{69.11}} & \bf{\color{Maroon!80}{47.62}} \\
				\cdashlinelr{1-12}
				\checkmark & & \checkmark & GR + GV$^\dagger$  & 17.433 & 371.8  & 26.01 & 23.96 & 65.29 & 38.49 & 71.00 & 48.27 \\
				\checkmark & & \checkmark & GR + RRT$^\dagger$ & 2.136  & 371.8  &  26.23 & 24.07 & 66.90 & 40.43 & 71.29 & 48.15 \\
				\checkmark & & \checkmark & \textbf{Ours}      & 0.345  & 387.4   & \bf{27.45} & \bf{26.06} & \bf{75.08} & \bf{49.81} & \bf{76.37} & \bf{57.53} \\
				\checkmark & & \checkmark & ~Ours$^{\ddagger}$               & \bf{0.345} & \bf{16.5} & \bf{\color{Maroon!80}{25.48}} & \bf{\color{Maroon!80}{23.82}} & \bf{\color{Maroon!80}{67.32}} & \bf{\color{Maroon!80}{43.80}} & \bf{\color{Maroon!80}{70.53}} & \bf{\color{Maroon!80}{48.75}} \\
				\bottomrule
			\end{tabular} 
		}
		\caption{Global retrieval followed by local feature re-ranking.}%
		\label{tab:global_local_agg}
	\end{subtable}
	\vspace*{-10pt}
	\caption{ \textbf{Applicability analysis}. $^\ddagger$: \emph{asymmetric retrieval setting}. $\downarrow$: lower is better; $^\dagger$: re-implementation; \Th{Ret}.: average retrieval latency for the $\gR$Oxf + 1M dataset. \Th{Ext}.: average feature extraction latency for the query side. \textbf{ASMK}$^*$~\cite{ASMK} aggregates local features into binarized embedding. \textbf{GR}: global retrieval; \textbf{GV}: Geometric Verification~\cite{oxford}; \textbf{RRT}: Re-ranking Transformers~\cite{RRT}. Re-ranking is performed on the \textbf{top}-$\bf{100}$ candidates returned by global retrieval. HOW~\cite{HOW} and $^\star$DELG~\cite{DELG} are two types of local feature with top performance. Mixer is adopted as gallery model. MobileNetv2 and Mixer are adopted as query model under \textbf{asymmetric} and \textbf{symmetric} setting, respectively.\vspace*{-15pt}}\label{tab:Applicability}
\end{table}

\begin{table}[tb]
	\centering
	\resizebox{1\linewidth}{!}
	{
		\small
		\setlength{\extrarowheight}{-1.5pt}
		\setlength{\tabcolsep}{2.0pt}
		\begin{tabular}{c cc cc cc}
			\toprule 
			\mr{2}{ \Th{Momentum} $\alpha$ } & \mc{2}{GLDv2-Test} & \multicolumn{2}{c}{$\gR$Oxf + 1M} & \multicolumn{2}{c}{$\gR$Par + 1M} \\
			
			\cmidrule(l){2-3} \cmidrule(l){4-5}  \cmidrule(l){6-7}
			
			& Private & Public & Medium & Hard & Medium & Hard \\
			
			\midrule
			0      & 25.86 & 24.29 & 62.75 & 43.19 & 71.46 & 50.63 \\
			0.5    & 27.42 & 24.71 & 66.36 & 45.69 & 75.18 & 55.75 \\
			0.9    & \bf{29.90} & 26.74 & 67.76 & 46.20 & 78.55 & 60.05 \\
			0.99   & 29.85 & \bf{27.68} & \bf{70.47} & \bf{49.16} & \bf{80.01} & \bf{62.58} \\
			0.999  & 29.81 & 27.17 & 69.52 & 48.62 & 79.04 & 61.28 \\
			\bottomrule
		\end{tabular} 
	}
	\vspace*{-8pt}
	\caption{ Analysis of $\bm{\alpha}$ \textbf{in Eq.~(\ref{equ:momentum_update})}. MobileNetV2 and mixer are adopted as query and gallery models, respectively.\vspace*{-8pt}}\label{tab:momentum}
\end{table}

\begin{table}[tb]
	\centering
	\resizebox{1\linewidth}{!}
	{
		\small
		\setlength{\extrarowheight}{-1.5pt}
		\setlength{\tabcolsep}{1.0pt}
		\begin{tabular}{cc cc cc cc}
			\toprule 
			\mr{2}{ \Th{Method}} & \mr{2}{ \tabincell{c}{ \Th{Two} \\ \Th{-stage} } } & \mc{2}{GLDv2-Test} & \multicolumn{2}{c}{$\gR$Oxf + 1M} & \multicolumn{2}{c}{$\gR$Par + 1M} \\
			
			\cmidrule(l){3-4} \cmidrule(l){5-6}  \cmidrule(l){7-8}
			
			& & Private & Public & Medium & Hard & Medium & Hard \\
			
			\midrule
			CSD~\cite{CSD} & $\checkmark$ & 25.51 & 23.73 & 64.42 & 43.90 & 68.32 & 47.76 \\
			Mixer + CSD  & $\checkmark$ & 27.23 & 24.73 & 67.08 & 45.50 & 76.69 & 57.74 \\
			Mixer + CSD  & $\times$ & \bf{28.73} & \bf{26.28} & \bf{69.66} & \bf{46.82} & \bf{78.41} & \bf{60.15} \\
			\cdashlinelr{1-8}
			LCE~\cite{LCE} & $\checkmark$ & 25.15 & 22.03 & 63.77 & 43.37 & 66.44 & 46.72 \\
			Mixer + LCE  & $\checkmark$ & 27.29 & 24.95 & 66.17 & 44.44 & 76.38 & 56.82 \\
			Mixer + LCE  & $\times$ & \bf{29.34} & \bf{27.35} & \bf{69.48} & \bf{48.95} & \bf{80.28} & \bf{62.70} \\
			\cdashlinelr{1-8}
			\mr{2}{\textbf{Ours}} & $\checkmark$ & 27.90 & 24.74 & 66.76 & 45.20 & 76.55 & 58.05 \\
			& $\times$ & \bf{29.85} & \bf{27.68} & \bf{70.47} & \bf{49.16} & \bf{80.01} & \bf{62.58} \\
			\bottomrule
		\end{tabular} 
	}
	\vspace*{-8pt}
	\caption{ Comparison of \textbf{two-stage and joint training (asymmetric retrieval)}. \Th{Two-satge}: mixer is first trained and kept fixed during the training of query model. MobileNetV2 and mixer are adopted as query and gallery models, respectively.\vspace*{-10pt}}\label{tab:two_or_one_stage}
\end{table}

\begin{table}[tb]
	\centering
	\resizebox{1\linewidth}{!}
	{
		\small
		\setlength{\extrarowheight}{-1.5pt}
		\setlength{\tabcolsep}{1.0pt}
		\begin{tabular}{lc cc cc cc}
			\toprule 
			\mr{2}{ \Th{Method} } & \mr{2}{\Th{Decoup.} } & \mc{2}{GLDv2-Test} & \multicolumn{2}{c}{$\gR$Oxf + 1M} & \multicolumn{2}{c}{$\gR$Par + 1M} \\
			
			\cmidrule(l){3-4} \cmidrule(l){5-6}  \cmidrule(l){7-8}
			
			& & Private & Public & Medium & Hard & Medium & Hard \\
			
			\midrule
			\mr{2}{ ~\tabincell{c}{ Mixer + \\ ~REG~\cite{AML} } }  & \texttimes & 12.61 & 11.74 & 46.29 & 26.55 & 49.57 & 31.00 \\
			& \checkmark &  \bf{15.87} & \bf{13.40} & \bf{49.91} & \bf{31.07} & \bf{52.25} & \bf{33.64} \\
			\cdashlinelr{1-8}
			\mr{2}{ \tabincell{c}{ Mixer + \\ ~LCE~\cite{LCE} } }  & \texttimes & 25.97 & 23.60 & 63.30 & 43.75 & 73.25 & 52.95 \\
			& \checkmark &  \bf{29.34} & \bf{27.35} & \bf{69.48} & \bf{48.95} & \bf{80.28} & \bf{62.70} \\
			\cdashlinelr{1-8}
			\mr{2}{ \tabincell{c}{ Mixer + \\ ~CSD~\cite{CSD} } }  & \texttimes & 18.13 & 16.42 & 48.31 & 29.49 & 47.13 & 29.94 \\
			& \checkmark & \bf{28.73} & \bf{26.28} & \bf{69.66} & \bf{46.82} & \bf{78.41} & \bf{60.15} \\
			\bottomrule
		\end{tabular} 
	}
	\vspace*{-0.5em}
	\caption{ \textbf{Generalization analysis}. \Th{Decoup.}: training processes of query model and mixer are decoupled with the momentum-updated mechanism. MobileNetV2 and mixer are adopted as query and gallery models, respectively.\vspace*{-10pt}}\label{tab:generalization}
\end{table}

\begin{table*}[tb]
	\centering
	\resizebox{1.0\linewidth}{!}{
		\normalsize
		\setlength{\extrarowheight}{-2.0pt}
		\setlength{\tabcolsep}{2.5pt}
		\begin{tabular}{l l cc cc cc cc cc cc cc}
			\toprule
			\mc{2}{\multirow{2}{*}{\Th{Method}}} & \mr{2}{\tabincell{c}{ \Th{Query} \\ \Th{Net} $\phi_q$}} & \mr{2}{\tabincell{c}{ \Th{Gallery} \\ \Th{Net} $\phi_g$}} & \mr{2}{\tabincell{c}{ \Th{Ext.} \\ (\emph{ms}) $\downarrow$}} & \mr{2}{\tabincell{c}{ \Th{Mem.} \\ (\emph{GB}) $\downarrow$}} & \multicolumn{2}{c}{GLDv2-Test} & \multicolumn{2}{c}{$\gR$Oxf} & \multicolumn{2}{c}{$\gR$Oxf + 1M} & \multicolumn{2}{c}{$\gR$Par}& \multicolumn{2}{c}{$\gR$Par + 1M} \\
			\cmidrule(lr){7-8}  \cmidrule(lr){9-10}  \cmidrule(lr){11-12} \cmidrule(lr){13-14} \cmidrule(l){15-16}
			& & & & & & Private & Public & Medium & Hard & Medium & Hard & Medium & Hard & Medium & Hard\\
			\midrule
			\multirow{16}{*}[-8pt]{\rotatebox[origin=c]{90}{\emph{\color{black!80}{Previous asymmetric image retrieval methods}}}} 
			
			&~~DELG$^\ddagger$\cite{DELG}     & R101      & R101  & 113.7 & 7.6 & 26.14 & 24.20 & 76.31 & 55.60 & 63.71 & 37.45 & 86.61 & 72.38 & 70.59  & 46.94 \\
			
			&~~~~~HVS~\cite{HVS}              &   \mr{3}{MobV2} & \mr{3}{R101} & 16.5 & 7.6 & 23.65 & 20.52 & 73.85 & 53.31 & 58.33  & 33.81 & 85.14 & 71.10 & 65.44  & 42.87 \\
			&~~~~~LCE~\cite{LCE}                 &                     &                     & 16.8 & 7.6& 23.88 & 20.45 & 73.95 & 52.78 & 57.76  & 32.45 & 84.76 & 70.78 & 66.22  & 43.03  \\
			&~~~~~CSD~\cite{CSD}                 &                     &                     & 16.5 & 7.6 & 23.33 & 21.33 & 74.72 & 53.70 & 60.19  & 34.14 & 85.70 & 71.46 & 68.25 & 44.24 \\
			
			\cmidrule(l){2-2}  \cmidrule(l){3-3}  \cmidrule(l){4-4} \cmidrule(l){5-5} \cmidrule(l){6-6}  \cmidrule(l){7-8}  \cmidrule(l){9-10} \cmidrule(l){11-12} \cmidrule(l){13-14} \cmidrule(l){15-16}
			
			&~~Token$^\ddagger$~\cite{Token}     & R101      & R101 & 131.1 & 3.8 & 29.20 & 27.24 & 82.16 & 65.75 & 70.58 & 47.46 & 89.40 & 78.44 & 77.24 & 56.81 \\
			&~~~~~HVS$^\dagger$~\cite{HVS}  & \mr{3}{MobV2} & \mr{3}{R101} & 16.3 & 3.8& 25.87 & 23.51 & 73.16 & 53.19 & 57.68 & 35.52 & 84.27 & 71.43 & 62.90 & 41.00 \\
			&~~~~~LCE$^\dagger$~\cite{LCE}     &                     &                     & 16.6 & 3.8& 26.06 & 24.42 & 74.82 & 56.40 & 61.57 & 39.43 & 84.09 & 71.88 & 65.70 & 43.86 \\
			&~~~~~CSD$^\dagger$~\cite{CSD}  &                     &                     & 16.3 & 3.8 & \bf{\color{gray}{26.58}} & \bf{\color{gray}{24.10}} & 75.52 & 56.83 & 63.46 & 39.01 & 84.50 & 70.73 & 65.93  & 43.77 \\
			\cmidrule(l){2-2}  \cmidrule(l){3-3}  \cmidrule(l){4-4} \cmidrule(l){5-5} \cmidrule(l){6-6}  \cmidrule(l){7-8}  \cmidrule(l){9-10} \cmidrule(l){11-12} \cmidrule(l){13-14} \cmidrule(l){15-16}
			&~~CVNet$^\ddagger$~\cite{CVNet}     & R101      & R101           & 113.7 & 7.6 & 30.71 & 28.73 & 80.01 & 62.83 & 74.25 & 54.56 & 90.18 & 79.01 & 80.82 & 62.74\\
			&~~~~~HVS$^\dagger$~\cite{HVS}                 &    \mr{3}{MobV2} & \mr{3}{R101}  & 16.5 & 7.6 & 26.50 & 23.49 & 74.90 & 55.17 & 62.32 & 42.66 & 84.92 & 71.62 & 67.13 & 46.80 \\
			&~~~~~LCE$^\dagger$~\cite{LCE}                 &                     &                     & 16.8 & 7.6 & 25.15 & 22.25 & 75.95 & 57.87 & 63.77 & 43.37 & 83.66 & 69.71 & 66.44 & 46.72 \\
			&~~~~~CSD$^\dagger$~\cite{CSD}                 &                     &                     & 16.5 & 7.6 & 25.51 & 23.73 & \bf{\color{gray}{76.44}} & \bf{\color{gray}{58.41}} & \bf{\color{gray}{64.42}} & \bf{\color{gray}{43.90}} & 85.32 & 71.61 & 68.32  & 47.76 \\
			\cmidrule(l){2-2}  \cmidrule(l){3-3}  \cmidrule(l){4-4} \cmidrule(l){5-5} \cmidrule(l){6-6}  \cmidrule(l){7-8}  \cmidrule(l){9-10} \cmidrule(l){11-12} \cmidrule(l){13-14} \cmidrule(l){15-16}
			&~~DOLG$^\ddagger$~\cite{DOLG}       & R101      & R101           & 126.7 & 1.9 & 29.46 & 26.85 & 82.37 & 64.94 & 75.19 & 53.55 & 90.97 & 81.71 & 82.28 & 66.45\\
			&~~~~~HVS$^\dagger$~\cite{HVS}                 &    \mr{3}{MobV2} & \mr{3}{R101}   & 16.1& 1.9 & 24.99 & 22.03 & 72.79 & 54.20 & 63.29 & 41.74 & 85.18 & 70.72 & 68.13 & 48.25 \\
			&~~~~~LCE$^\dagger$~\cite{LCE}                 &                     &                     & 16.3 & 1.9 & 26.04 & 24.06 & 72.84 & 53.70 & 61.90 & 40.84 & 85.77 & 69.54 &67.65 & 48.53 \\
			&~~~~~CSD$^\dagger$~\cite{CSD}                 &                     &                     & 16.1 & 1.9 & 25.64& 24.04 & 75.53 & 56.23 & 64.02 & 42.79 & \bf{\color{gray}{86.34}} & \bf{\color{gray}{72.84}} & \bf{\color{gray}{69.29}}  & \bf{\color{gray}{49.47}} \\

			
			\midrule
			\multirow{7}{*}[-1pt]{\rotatebox[origin=c]{90}{\emph{\color{black!80}{Feature Fusion}}}}
			&~~Ensemble$^{5,632}$  & - & - & 485.3 & 20.9 & 32.28 & 29.79 & 82.86 & 66.06 & 75.30 & 53.24 & 90.74 & 81.72 & 83.35 & 67.05\\
			&~~~~~\textbf{Ours$^{512}$}       & Mixer      & Mixer           & 492.7 & 1.9 & \bf{32.93} & \bf{31.63} & 85.16 & 70.35 & 77.58 & 58.30 & 91.38 & 82.41 & 83.68 & 68.04\\
			&~~~~~\textbf{Ours$^{512}$}       & MobV2      & Mixer     & 16.1 & 1.9 & \color{Maroon!80}{28.57} & \color{Maroon!80}{26.57} & \color{Maroon!80}{79.46} & \bf{\color{Maroon!80}{64.18}} & \color{Maroon!80}{68.17} & \color{Maroon!80}{47.27} & \color{Maroon!80}{90.44} & \color{Maroon!80}{79.64} & \color{Maroon!80}{77.90} & \color{Maroon!80}{59.01}\\
			\multicolumn{6}{c}{~~~~~\textbf{\emph{\color{gray}{mAP gains over the previous asymmetric SOTA}}}} & \bIP{1.99} & \bIP{2.47} & \bIP{3.02} & \bI{5.77} & \bIP{3.75} & \bIP{3.37} & \bIP{4.10} & \bIP{6.80} & \bIP{8.61}  & \bIP{9.54} \\
			&~~~~~\textbf{Ours$^{2,048}$}       & Mixer      & Mixer           & 493.5 & 7.6 & 32.85 & 31.27 & \bf{85.24} & \bf{70.43} & \bf{77.84} & \bf{58.91} & \bf{91.55} & \bf{82.55} & \bf{84.43} & \bf{69.44}\\
			&~~~~~\textbf{Ours$^{2,048}$}       & MobV2      & Mixer     & 16.5 & 7.6 & \bf{\color{Maroon!80}{29.85}} & \bf{\color{Maroon!80}{27.68}} & \bf{\color{Maroon!80}{80.19}} & \color{Maroon!80}{64.14} & \bf{\color{Maroon!80}{70.47}} & \bf{\color{Maroon!80}{49.16}} & \bf{\color{Maroon!80}{90.48}} & \bf{\color{Maroon!80}{80.55}} & \bf{\color{Maroon!80}{80.01}} & \bf{\color{Maroon!80}{62.58}}\\
			\multicolumn{6}{c}{~~~~~\textbf{\emph{\color{gray}{mAP gains over the previous asymmetric SOTA}}}} & \bI{3.27} & \bI{3.58} & \bI{3.75} & \bIP{5.73} & \bI{6.05} & \bI{5.26} & \bI{4.14} & \bI{7.71} & \bI{10.72}  & \bI{13.11} \\
			\bottomrule
	\end{tabular}	}
	\vspace*{-5pt}
	\caption{\textbf{Comparison to the state-of-the-art methods.} $^d$: feature dimension is $d$; $\downarrow$: lower is better; $^\ddagger$: re-evaluate official public weights; $^\dagger$: re-implementation; \Th{Ext.}: latency of feature extraction for query side; \Th{Mem.}: average memory footprint for $\gR$1M dataset; MobV2: MobileNetv2; R101: ResNet101; Mixer: aggregate various features into a compact vector with $\phi^{\rm mix}$; Ensemble: feature concatenation.\vspace*{-10pt}}
	\label{tab:sota}
\end{table*}

\noindent \textbf{Analysis of the training strategy}. Our AFF adopts the momentum update mechanism to jointly train the mixer and query model. Here, we expect to answer two questions: 

(1) \emph{Whether momentum update is necessary or not?} Tab.~\ref{tab:momentum} shows the accuracy of the asymmetric retrieval with different momentum values $\alpha$ in Eq.~(\ref{equ:momentum_update}). The accuracy increases gradually as $\alpha$ increases. It performs well when $\alpha$ is in $0.99 \sim 0.999$, showing that a slowly progressing classifier $\psi_q$ is beneficial. The importance of the momentum update is also confirmed by the results in Tab.~\ref{tab:generalization}. When our AFF is combined with different approaches, coupling the training process of mixer and query model leads to significant performance degradation, \eg, the mAP on $\gR$Oxf + 1M drops from $69.66$ to $48.31$ for ``Mixer $+$ CSD~\cite{CSD}''. All these results support our motivation of decoupling the training of the mixer and query model. 

(2) \emph{Is joint training more effective than two-stage training?} Two-stage training means that the mixer is first trained, followed by the compatibility training of the query model. It avoids the difficulties arising in joint training but leads to a more complex and time-consuming training process. Besides, due to the limited capacity of the query model, directly aligning it to a static powerful model limits the feature compatible training, which is confirmed by previous methods~\cite{jin2019knowledge,mirzadeh2020improved}. Experiment results are shown in Tab.~\ref{tab:two_or_one_stage}. When our AFF is combined with various existing methods, joint training yields a consistent performance boost.

\noindent \textbf{Generalizability analysis}. In fact, our AFF is feasible to be combined with various existing methods, which leads to different training loss $\ell_{\rm comp.}(\phi_q, x)$. For example, when combined with REG~\cite{AML}, we maintain a momentum-updated version $\widetilde{\phi}_{\rm mix}$ of mixer $\phi_{\rm mix}$, whose output feature is denoted as $\widetilde{\bm{g}}^{\rm mix}$. Then, $\ell_{\rm comp.}(\phi_q, x)$ is formulated as $\ell_{\rm comp.}(\phi_q, x) = - \ltnorm{\bm{q} -\widetilde{\bm{g}}^{\rm mix}}^2$. More details about the combinations with existing methods, \eg, CSD~\cite{CSD} and LCE~\cite{LCE}, are provided in the supplementary materials. As shown in Tab.~\ref{tab:generalization}, our method enhances all existing asymmetric retrieval systems, leading to a promising accuracy improvement without adding any overhead to the query side. All the results demonstrate the generalizability of our method.

\subsection{Comparison to the state-of-the-art methods}
\label{sub:sota} 
In Tab.~\ref{tab:sota}, our method is compared with the state-of-the-art asymmetric methods (SOTA). Previous asymmetric retrieval methods, adopting a single feature on the gallery side, still suffer severe performance degradation especially in the large-scale case. Our method performs feature fusion at the gallery side, which significantly improves the accuracy of asymmetric retrieval, \eg, the mAP on $\gR$Oxf + 1M increases from $64.42$ to $70.47$. Besides, compared to direct feature ensemble, our method introduces no extra computation and storage overhead to the query side, which is friendly for various resource-constrained platforms, \eg, the latency of feature extraction on the query side is comparable to the previous asymmetric retrieval method, \eg, $16.5$ \emph{ms} for ours \vs $16.1$ \emph{ms} for previous SOTA.

\section{Conclusion}
\label{sec:conclusions}
In this work, we introduce a new asymmetric feature fusion paradigm to enhance the accuracy of existing asymmetric systems. It alleviates the dilemma of trade-off between retrieval efficiency and accuracy, caused by the limited capacity of lightweight models. The proposed paradigm deploys various models at the gallery side to extract features, which are further aggregated into compact embedding with a dynamic mixer for efficient retrieval. As for the query side, only a single lightweight model is deployed for feature extraction. The query model and the mixer are jointly trained to achieve feature compatibility. With the proposed framework, the accuracy of asymmetric retrieval is significantly boosted without introducing any additional computational and storage overhead to the query side. Experiments on several datasets demonstrate the wide applicability and excellent effectiveness of our paradigm. 

\footnotesize {\flushleft \bf Acknowledgements}. This work was supported in part by the National Natural Science Foundation of China under Contract 62102128 and 62021001, and in part by the Fundamental Research Funds for the Central Universities under contract WK3490000007. It was also supported by the GPU cluster built by MCC Lab of Information Science and Technology Institution and the Supercomputing Center of the USTC.

{\small
	\bibliographystyle{ieee_fullname}
	\bibliography{biblio}
}
\end{document}